\documentclass[sn-mathphys,Numbered]{sn-jnl}

\usepackage{graphicx}%
\usepackage{multirow}%
\usepackage{amsmath,amssymb,amsfonts}%
\usepackage{amsthm}%
\usepackage{mathrsfs}%
\usepackage[title]{appendix}%
\usepackage{xcolor}%
\usepackage{textcomp}%
\usepackage{manyfoot}%
\usepackage{booktabs}%
\usepackage{algorithm}%
\usepackage{algorithmicx}%
\usepackage{algpseudocode}%
\usepackage{listings}%

\usepackage{bm}%

\usepackage{macros}
\newcommand{\act}{\nu} 
\newcommand{\ab}{_{\alpha\beta}} 
\newcommand{\kcut}{k_{\rm cut}}
\newcommand{\elektrum}{\textit{Elektrum}}

\newcommand{\revision}[1]{\textcolor{black}{#1}}

\theoremstyle{thmstyleone}%
\theoremstyle{thmstyletwo}%

\theoremstyle{thmstylethree}%

\begin{document}

\title[Article Title]{Interpretable neural architecture search and transfer learning for understanding \revision{CRISPR/Cas9 off-target} enzymatic reactions}


\title[Article Title]{Interpretable neural architecture search and transfer learning for understanding \revision{CRISPR/Cas9 off-target} enzymatic reactions}

\author[1]{\fnm{Zijun} \sur{Zhang}}\email{zijun.zhang@cshs.org}
\equalcont{These authors contributed equally to this work.}

\author[2]{\fnm{Adam R.} \sur{Lamson}}\email{alamson@flatironinstitute.org}
\equalcont{These authors contributed equally to this work.}

\author*[2,3]{\fnm{Michael} \sur{Shelley}}\email{mshelley@flatironinstitute.org}

\author*[2,4]{\fnm{Olga} \sur{Troyanskaya}}\email{ogt@genomics.princeton.edu}

\affil[1]{\orgdiv{Division of Artificial Intelligence in Medicine }, \orgname{Cedars-Sinai Medical Center}, \orgaddress{\street{116 N. Robertson Blvd}, \city{Los Angeles}, \postcode{90048}, \state{CA}, \country{USA}}}

\affil[2]{\orgdiv{Center for Computational Biology}, \orgname{Flatiron Institute}, \orgaddress{\street{162 5th Ave}, \city{New York City}, \postcode{10010}, \state{NY}, \country{USA}}}

\affil[3]{\orgdiv{Courant Institute of Mathematical Sciences}, \orgname{New York University}, \orgaddress{\street{251 Mercer Street}, \city{New York City}, \postcode{10012}, \state{NY}, \country{USA}}}

\affil[4]{\orgdiv{Lewis Sigler Institute for Integrative Genomics}, \orgname{Princeton University}, \orgaddress{\street{Carl Icahn Laboratory
      South Drive}, \city{Princeton}, \postcode{08544}, \state{NJ}, \country{USA}}}


\abstract{Finely-tuned enzymatic pathways control cellular processes, and their dysregulation can lead to disease. Creating predictive and interpretable models for these pathways is challenging because of the complexity of the pathways and of the cellular and genomic contexts. Here we introduce \textit{Elektrum}, a deep learning framework which addresses these challenges with  data-driven and biophysically interpretable models for determining the kinetics of biochemical systems. First, it uses \textit{in vitro} kinetic assays to rapidly hypothesize an ensemble of high-quality Kinetically Interpretable Neural Networks (KINNs) that predict reaction rates. It then employs a novel transfer learning step, where the KINNs are inserted as intermediary layers into deeper convolutional neural networks, fine-tuning the predictions for reaction-dependent \textit{in vivo} outcomes. \textit{Elektrum} makes effective use of the limited, but clean \textit{in vitro} data and the \revision{complex}, yet plentiful \textit{in vivo} data that captures cellular context. We apply \textit{Elektrum} to predict CRISPR-Cas9 off-target editing probabilities and demonstrate that \textit{Elektrum} achieves state-of-the-art performance, regularizes neural network architectures, and maintains physical interpretability.
}

\keywords{Interpretable neural networks, transfer learning, genome editing, AutoML, neural architecture search}



\maketitle

\section{Introduction}
\label{sec:introduction}

Enzymatic reactions control the rate at which cellular processes occur. Proteins are broken down, energy is stored, and DNA is replicated thanks to the finely tuned enzymatic pathways of our cells. Dysregulation of these pathways can lead to various diseases, making the study of them an important part in biomedical research \cite{gebauer2021rna, masoud2015hif,  santamaria2020adamts, flinn2018adenosine}.
However, these reaction pathways can be extremely complex, traversing many intermediate states with unknown kinetic rates before obtaining a final product or event \cite{kim2019kinetic, persikov2015systematic,Liepelt2007,Schreiber2002}.
This makes discovering and testing biochemical kinetics models an arduous task. This task is further complicated by the fact that the cellular environment is \revision{complex} and introduces confounding factors. To better understand the chemical kinetics of our cells, we require new analysis tools.

Deep learning is now a commonly used tool in the biological sciences, being applied to tasks such as analyzing images, predicting protein structures, and processing genomic data. Neural networks (NNs) are especially effective at finding patterns in genomic datasets \cite{Zhou2015, fudenberg2020predicting, Li2021b, avsec2021effective, wong2021decoding, ching2018opportunities}.
This is because NNs train efficiently with gradient descent and yet can have great predictive power with little prior knowledge. However, given NNs’ generic structure and large number of trained parameters, it is difficult to gain deeper scientific insights \revision{or generalize to systems outside of the datasets on which they were trained}.

Interpretable neural networks (INNs), on the other hand, incorporate prior knowledge and/or can be examined to extract more information on the systems they model. Examples of INN include free-energy architectures that map exactly to a partition function-like formula. Upon training, the network node parameters are the chemical reactions’ \revision{sequence-dependent} equilibrium constants. Free-energy neural networks have been applied to phenotypic expression assays and deep-mutational protein scanning assays \cite{Tareen2019, Tareen2022, Fowler2014, Faure2022}. A less explored, yet more informative INN architecture is the \revision{Kinetically} Interpretable Neural Network (KINN). KINNs correspond to a set of linear ordinary differential kinetic equations for the relative occupation of enzymatic states, and whose coefficients are state transition rate constants \cite{Tareen2019}. These learned rates allow prediction of more complicated biochemical scenarios, like non-steady-state \revision{systems and multiple species reactions}, by calculating the kinetic system's eigenvalues(in preparation).

Unfortunately, KINNs and free-energy neural networks suffer several limitations, restricting wide adoptions. First, they require a candidate kinetic model before training, losing the advantage of needing minimal information when creating the machine learning model. Furthermore, training datasets are restricted to data generated by specific \textit{in vitro} assays \cite{Kretz2015, Persikov2015, Jones2021}. While \textit{in vitro} data harbors little noise or confounding effects compared to their \textit{in vivo} counterparts, it is challenging to evaluate how well \textit{in vitro} results and predictions apply to \textit{in vivo} scenarios. Finally, \textit{in vitro} kinetic and free-energy datasets are less common and smaller, making it harder to train and validate machine learning model results. Even though \textit{in vitro} trained models have shown promise, these challenges undermine their performance when applied to understanding \textit{in vivo} settings, which is the ultimate goal.

We introduce a new deep learning framework called \textit{Elektrum} that generates data-driven mechanistically interpretable networks for \textit{in vivo} kinetic systems by integrating \revision{data from} \textit{in vitro} and \textit{in vivo} assays. One of \textit{Elektrum}’s key innovations is its transfer learning step where instances of a kinetic model ensemble are tested as intermediate layers of a deeper convolutional neural network (CNN). This leverages the power of transfer learning that repurposed a trained model architecture to a new but related task, enabling effective use of both the limited but clean \textit{in vitro} data and plentiful but \revision{complex} \textit{in vivo} data. Our transfer learning method is distinct from conventional transfer learning because we train a smaller NN on a specialized dataset to achieve interpretability, then gain deeper insight by augmenting the network with a context-aware CNN backbone.

We show this approach regularizes the NN architecture while maintaining interpretability, even for the ultimate challenge of predicting \textit{in vivo} processes.  We applied \textit{Elektrum} to model CRISPR-Cas9 off-target editing kinetics. Existing CRISPR/Cas9 predictors fall into one of two classes: accurate, data-driven machine learning models or expert-derived, mechanistic biophysical models. Our model is the first to overcome the  trade-off of predictability versus mechanistic interpretation by achieving both state-of-the-art performance for predicting \textit{in vivo} off-target editing and providing physical interpretability of the enzymatic kinetic scheme.

\begin{figure}[h]
  \centering
  \includegraphics[width=0.9\textwidth]{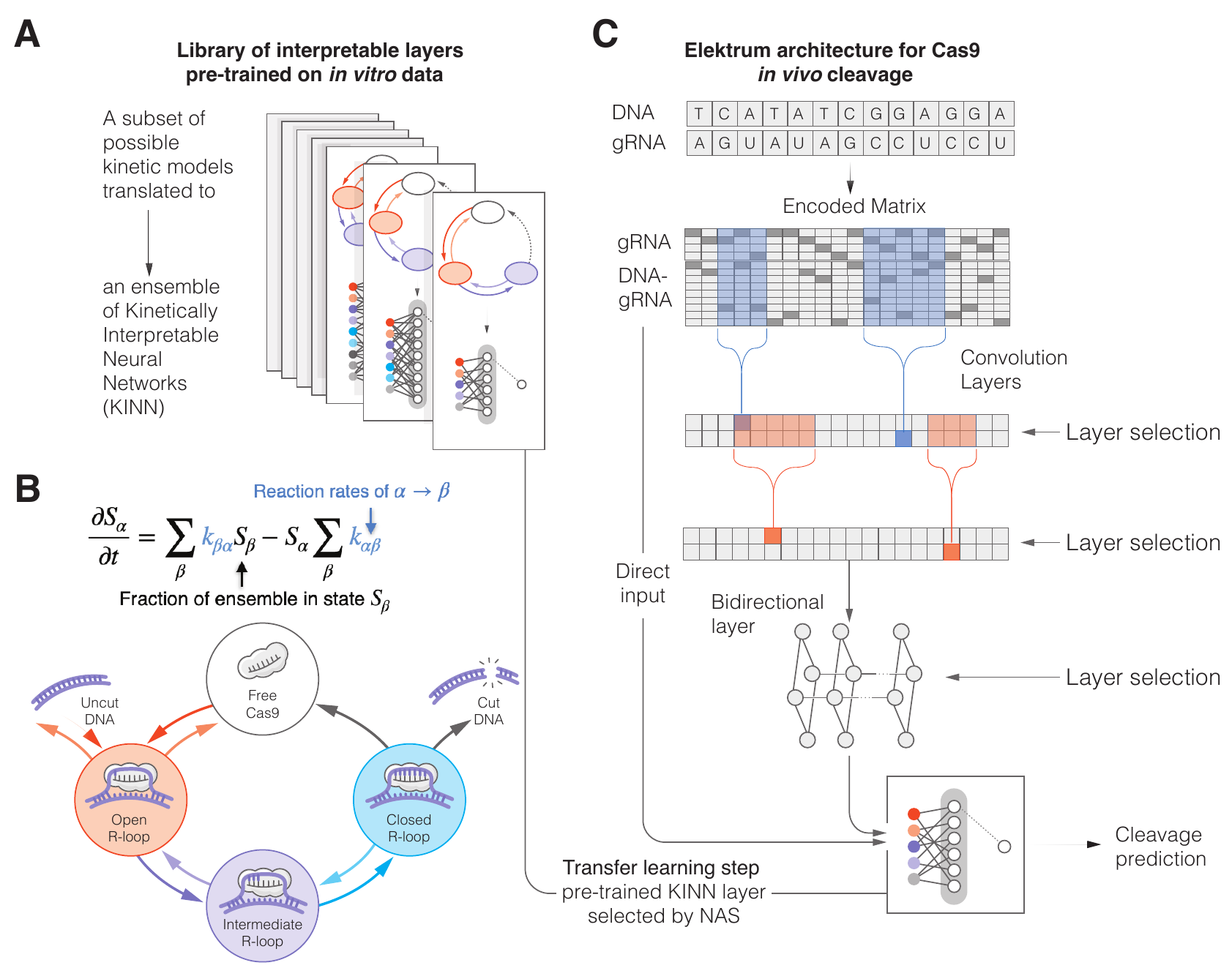}
  \caption{Overview of \textit{Elektrum} framework. A) We use a probabilistic model-building genetic algorithm to search for and create a set of \revision{Kinetically Interpretable Neural Networks} (KINNs), a special type of sparse neural networks that has one-to-one correspondence with kinetic models. These KINNs differ in number of enzymatic states and reaction rates but model the \textit{in vitro} data equally well. B) KINNs can be translated to a system of first-order rate equations. For CRISPR-Cas9, this corresponds to a multi-step enzymatic reaction, where the kinetic rate for each enzymatic step is determined by the DNA-gRNA sequence pair. C) Transfer learning integrates KINNs pretrained on \textit{in vitro} data with a deep neural network to predict \textit{in vivo} Cas9 cleavage. We employ neural architecture search (NAS) to optimize the convolutional and bidirectional layer while searching for an optimal static KINN layer. This integrative approach expands the types and amount of applicable training datasets by many fold while maintaining kinetic interpretability.
  }
  \label{fig:schematic}
\end{figure}

\section{Results}
\label{sec:results}

\subsection{Overview of \textit{Elektrum} Framework}
\label{sub:overview}

\textit{Elektrum} is a deep learning framework for modeling \textit{in vitro} and \textit{in vivo} sequence-dependent enzymatic kinetics (Figure \ref{fig:schematic}). By utilizing neural architecture search (NAS) \cite{Zhang2021AMBER} and transfer learning techniques, \textit{Elektrum} automatically discovers a system’s underlying kinetics by generating and testing different KINN architectures. While searching, \textit{Elektrum} creates a set of candidate KINNs that predict a measurable kinetic rate \textit{in vitro}. Leveraging the resulting ensemble of possible kinetic models, we then employ a second transfer learning NAS step to search for a hybrid CNN+KINN model. This second step extends the KINN architecture by using a more complex CNN backbone to learn nuanced sequence-dependent signals hidden in the \textit{in vivo} data. The NAS incorporates pretrained KINN \revision{models} as an intermediate layer, selecting for one of the KINNs that maximizes \textit{in vivo} model performance.

When searching the KINN architecture space, \textit{Elektrum} efficiently builds multiple candidate kinetic models in a data-driven fashion, in contrast to a single kinetic system derived by human experts. Briefly, \textit{Elektrum} defines a set of prior distributions over the KINN model architecture and updates the posterior probabilities based on performance (Methods). A candidate KINN’s convolutional first layer samples DNA-gRNA subsequences to model the reaction-rate/sequence dependency. \textit{Elektrum} then gathers the kinetic rates between enzymatic states and employs a 3-layer King-Altman neural network to calculate the steady-state production rate of an experimentally measurable quantity, such as cleavage products. Alternatively, one may use a differentiable eigen-decomposition layer to model non-steady-state kinetic system dynamics, which is important for understanding protein binding and cleavage in experimental settings (Methods). The KINN’s performance on a held-out dataset is used to update the posterior model architecture distributions following a Bayesian probabilistic genetic algorithm (Methods). When benchmarked on simulated kinetic systems under a variety of different settings, we show that our search algorithm identified substantially more accurate models compared to random sampling (Extended Data Figure 1). \revision{Importantly, KINNs have the one-to-one correspondence with a set of linear ordinary differential equations (ODEs), therefore enabling direct discovery of biophysical knowledge. This is in contrast to conventional model interpretation and feature attribution methods, where deriving a set of kinetic ODEs from the feature importance is non-trivial.}

Given a set of pretrained KINNs, it is crucial to evaluate how these \textit{in vitro} kinetic models apply to \textit{in vivo} reactions. This evaluation is challenging because \textit{in vivo} enzymatic processes are more complex, and may not have directly measurable products. Two innovations make this evaluation possible. First, pretrained KINNs are used as special neural network layers, such that \textit{in vitro} kinetic systems with different numbers of states and kinetic rates are searched to optimize the \textit{in vivo} model’s performance. Second, \textit{Elektrum} accounts for perturbative effects to the kinetic rates from \textit{in vivo} sequence context by adding a deep CNN backbone. This integrative approach expands the types and amount of applicable training datasets by many fold while maintaining kinetic interpretability. As we demonstrate below, the transfer learning scheme substantially improves the final model performance.

\begin{figure}[h]
  \centering
  \includegraphics[width=0.9\textwidth]{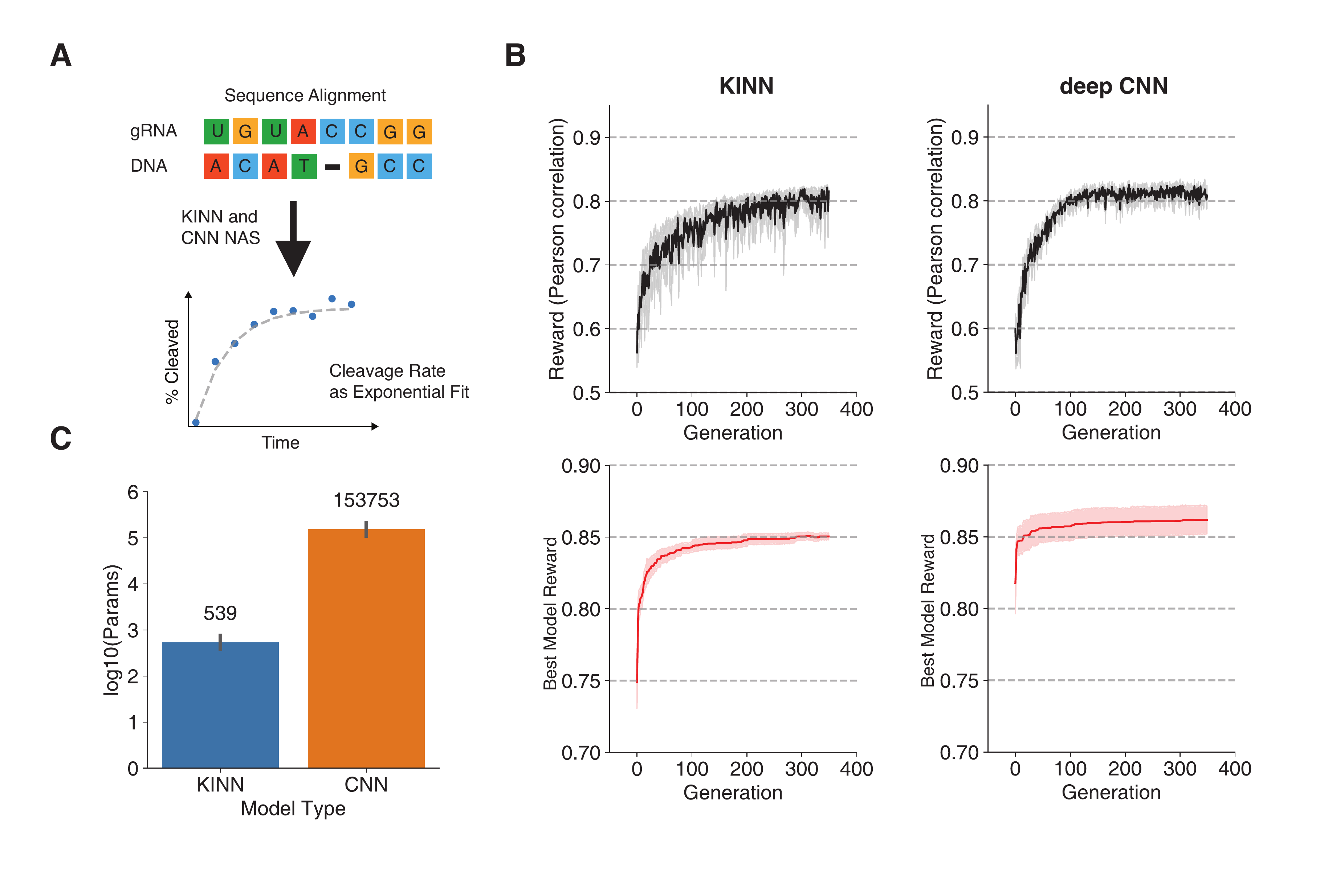}
  \caption{\revision{Searching} KINN \revision{architectures} for \textit{in vitro} Cas9 cleavage kinetics. A) Schematic of the machine learning task for predicting Cas9 cleavage kinetics. A model takes a gRNA-DNA alignment as input to predict the cleavage rate by fitting a single exponential equation in massively parallel kinetic profiles. We trained five replicates of KINNs and deep CNN models, optimizing both of the model architectures to maximize the accuracy of DNA cleavage rate. B) Search performance \revision{dynamics} for KINN and deep CNN across \revision{NAS} iterations (black lines)\revision{, as measured by Pearson correlation in held-out validation data}. \revision{We further visualized the Pearson correlation reward for the best individual model found so far at each generation (red lines), demonstrating better models were found over time}. Shaded areas are 95\% confidence intervals. C) Number of trainable parameters between the optimized KINN and optimized CNNs. Error bars represent standard deviations of best models across five independent search runs.}
  \label{fig:kinn_perf}
\end{figure}

\subsection{Modeling \textit{in vitro} Cas9 off-target cleavage by KINN search}
\label{sub:kinn_search}

\revision{W}e apply \textit{Elektrum} to model the enzymatic pathway of the Cas9 protein as it binds and cleaves a DNA sequence (Figure \ref{fig:schematic}B). When charged with a 20 nucleotide (nt) guide RNA molecule (gRNA), Cas9 enzymes will attach to a complementary DNA target that it subsequently cleaves. However, partial sequence matches may also be cleaved, though typically at lower rates, which are demonstrably detrimental to the safety of gene editing therapies \cite{tsai2017circle, haapaniemi2018crispr, cancellieri2023human}.
Therefore, the cleavage rate dependence on gRNA-DNA match poses an important kinetic learning problem for off-target editing \cite{Eslami-Mossallam2022, Klein2018, Fu, concordet2018crispor, lin2020crispr}.
To learn Cas9 cleavage kinetics, we utilized the only published \textit{in vitro} Massively Parallel Kinetic Profiling (MPKP) dataset of Cas9 cleavage\cite{Jones2021}. Two gRNAs were profiled and their time-resolved concentrations of cut DNA molecules were measured from a library of substrates with saturated mutations and indels. This experiment measured their absolute cleavage rates by fitting a single exponential rate function for each gRNA-DNA pair.

\textit{Elektrum} optimizes KINNs to accurately predict Cas9 cleavage rates by first creating a KINN model space – the number of intermediate kinetic rates, rate dependencies, and enzymatic states – using an expert-curated kinetic model \cite{Klein2018,Eslami-Mossallam2022} as a weak prior (Methods). \revision{\textit{Elektrum} leverages a genetic algorithm to sample KINNs from a model space with prior probability centered around expert knowledge of a four-state kinetic model reported previously \cite{Eslami-Mossallam2022}. Note that this is different from KINNs being initiated from the expert-curated kinetic model.}
The model's predictive power improved as \textit{Elektrum} iteratively optimized the KINN architectures (Figure \ref{fig:kinn_perf}B).  \textit{Elektrum}-optimized KINN performed 25.2\% better in test Pearson correlation than the baseline randomly sampled CNNs on the same feature and label set \revision{(Extended Data Figure 1)}.
\revision{Our searched KINNs were highly consistent with the expert-curated model (Extended Data Table 1).}
More impressively, these KINNs perform on par with the state-of-the-art accuracy of non-interpretable deep CNNs identified by neural architecture search \cite{Zhang2021AMBER} \revision{(Figure \ref{fig:kinn_perf}B)}. And, besides providing interpretability, the optimized KINNs require a factor of $2.4\times10^2$ fewer parameters compared to traditional CNNs (Figure \ref{fig:kinn_perf}C).
\revision{This reduction in parameters also translates to faster training time compared to traditional CNNs, especially for larger batch sizes (Extended Data Figure 2).}

We explored different candidate high-quality kinetic systems with different numbers of enzymatic states. To integrate multiple different kinetic models, we analyzed the architectures’ posterior distributions using a Bayesian hierarchical multivariate approach, thus retaining the variability between models (Methods). Given a fixed number of states, the model architectures discovered by \textit{Elektrum} were highly consistent between the gRNA1 and gRNA2 datasets (Extended Data Figure \revision{3}).
For both datasets, when exploring more complex models, increasing the number of states from 4 to 5 gives a splitting of the first enzymatic state into two sequential states. Similarly, when going from 5 to 6 states, the last state is split into two sequential states. This suggests that \textit{Elektrum} is discovering more complex free-energy landscapes that are being coarse-grained at fewer states, especially at the reaction’s first and last stages. All models with different numbers of states (4, 5, or 6) performed comparably well on the \textit{in vitro} data (Extended Data Table \revision{2}).
This exploration of state number constructs a natural model ensemble for our \textit{in vivo} transfer learning step, from which one KINN architecture \revision{and its pretrained kinetics weights} will be selected based on the sequence and cellular context provided by \textit{in vivo} data.

\subsection{\textit{in vitro} KINN models partially capture \textit{in vivo} off-target editing}
\label{sub:kinn_off-target}

Precise prediction of off-target cleavages is critical for effective Cas9 deployment for basic research and safe therapeutic use in humans. Many experimental approaches, such as GUIDE-Seq, CIRCLE-Seq, SITE-Seq \cite{tsai2017circle, listgarten2018prediction, doench2016optimized, haeussler2016evaluation, cameron2017mapping, kleinstiver2015engineered},
are designed to measure cleaved off-target sites in living cells for specific time periods and concentrations of gRNA-charged Cas9 RNP. These data are key for developing biological insight and for developing predictive methods, but on their own do not provide a comprehensive framework for predicting off-target cleavage frequency and time dependencies. Here, we use these \textit{in vivo} test datasets as holdouts to evaluate \textit{in vivo} performance of KINNs trained on \textit{in vitro} data to predict Cas9 off-targeting cleavage rates.

\begin{table}[h]
  \begin{tabular*}{\textwidth}{@{\extracolsep\fill}lccc}
    \toprule%
    & \multicolumn{2}{@{}c@{}}{Mutation-Only Data}
    & {Mutation + Indel Data} \\
    \cmidrule{2-3}\cmidrule{4-4}%
    Method & AUPR(Kleinstiver) & AUPR(Listgarten) & AUPR(Listgarten)  \\
    \midrule
    \textit{Elektrum}  & \bf{0.364} & 0.262 & \bf{0.324} \\
    KINN & 0.202 & 0.079  & 0.103\\
    AMBER-CNN & 0.145 & 0.081  & 0.060\\
    CRISPR-Net \cite{lin2020crispr} & 0.329 & \bf{0.317}  & 0.254\\
    AttnToMismatch\_CNN \cite{liu2019prediction} & 0.071 & 0.025  & $-$\\
    Elevation-score \cite{listgarten2018prediction} & 0.131 & 0.078  & $-$\\
    CFD \cite{doench2016optimized} & 0.066 & 0.030  & $-$\\
    Ensemble SVM \cite{Peng2018} & 0.113 & 0.048  & $-$\\
    CNN\_std \cite{lin2018off} & 0.115 & 0.034  & $-$\\
    CRISPRoff \cite{alkan2018crispr} & 0.104 & 0.046  & $-$\\
    \revision{Baseline} & \revision{0.00056} & \revision{0.00014}  & \revision{0.00028} \\
    \botrule
  \end{tabular*}
  \caption{External \revision{\textit{in vivo}} data performance comparisons. In this work we proposed three new methods: KINN is an architecture-optimized kinetic model using \textit{in vitro} MPKP data, while AMBER-CNN is an architecture-optimized deep CNN model trained with the same feature and label set as KINN. \textit{Elektrum} is a hybrid KINN+CNN optimized model, trained with our novel transfer learning strategy. \revision{Existing machine learning methods are trained using various training datasets and are detailed in the corresponding publications. Baseline of AUPR is the proportion of positive labels in a test dataset, the expected AUPR value from a random classifier.}}\label{tab:performance}
  \label{tab:compare}
\end{table}

We assessed the sequence cleavage predictions by comparing predicted on- and off-target cleavage rates to non-edited sequences. First, we found a strong positive correlation between the predicted cleavage rate and off-target editing levels (as indicated by the observed high-throughput sequencing read coverage; Extended Data Figure \revision{4}).
The KINN model accurately predicted the cleavage probability of off-target sites in \textit{in vivo} data when compared to conventional state-of-the-art Cas9 off-target predictors; only CRISPR-Net, which leveraged a 158 times larger training dataset (training data size 1,112,198 for CRISPR-Net, 6,976 for KINN), outperformed the KINN model (Table \ref{tab:performance}). When trained on the same \revision{\textit{in vitro}} data, the KINN’s \textit{in vivo} predictions even outperformed CNNs despite KINN’s slightly worse performance on the \textit{in vitro} data (Figure \ref{fig:kinn_perf}). This suggests that in the presence of data extrapolation and domain shifts, like the \textit{in vitro} training shifted to \textit{in vivo} testing, a mechanistic model could be more robust than a complex neural network.

Furthermore, by having a learned underlying mechanistic model, KINNs can be generalized to systems outside of the parameters inherent in the training data, providing a framework to design improved protocols for genome wide editing. Conventional predictors, on the other hand, are limited to the specific \textit{in vivo} experimental parameters in the training data which are at a fixed time exposure and concentration of Cas9. Thus these predictors are unable to model temporal dynamics of Cas9 activity, which is important for understanding and controlling editing specificity \cite{Zhuo2021}.

\begin{figure}[h]
  \centering
  \includegraphics[width=0.9\textwidth]{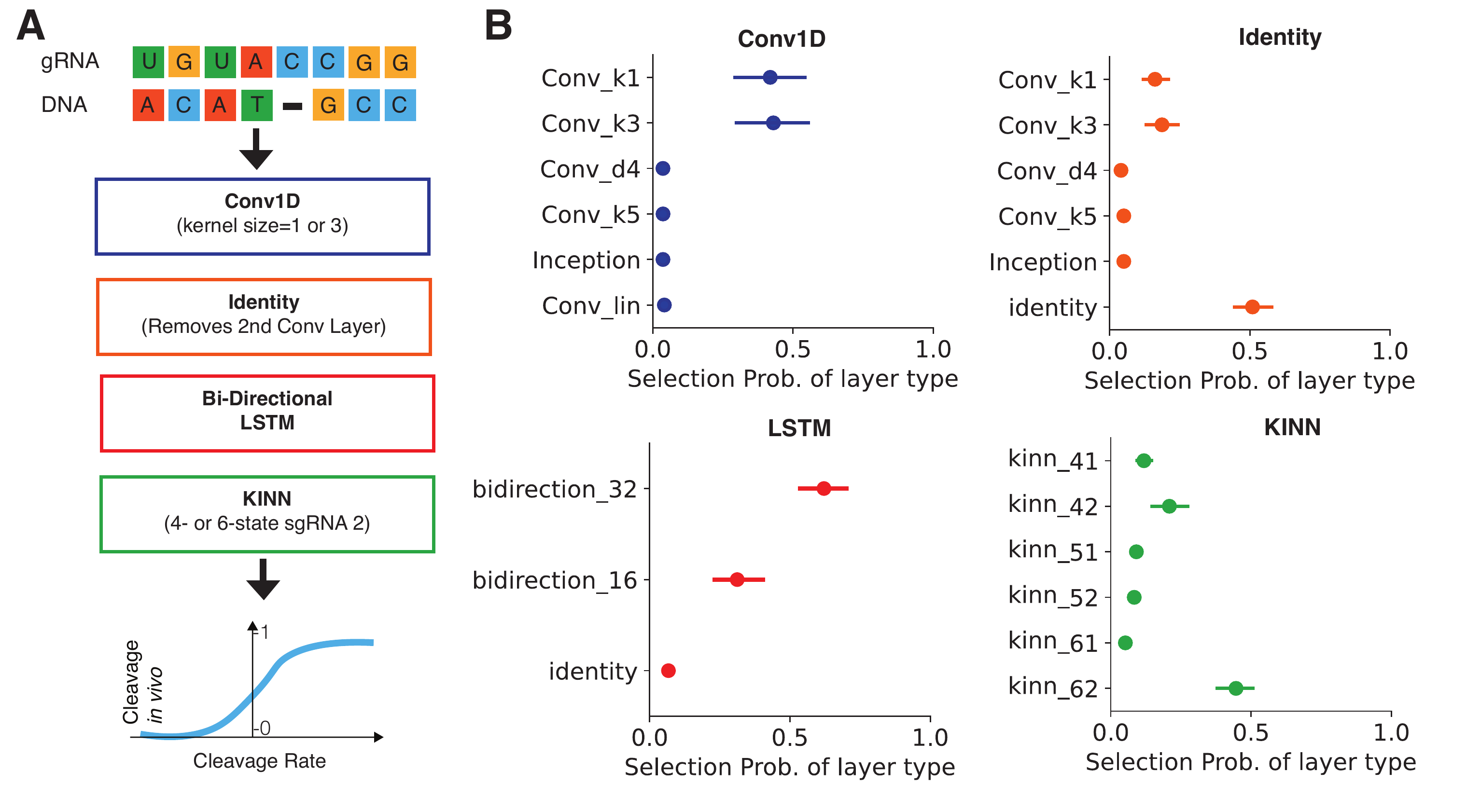}
  \caption{Integration of transfer learning in NAS for \textit{in vivo} cleavage prediction. A) Optimal deep learning model architecture. The model consists of a convolution layer with small kernel sizes, a complex bidirectional recurrent layer, and a KINN layer. The output of KINN layer’s cleavage rate is mapped to a binary label of edited vs unedited off-target sites \textit{in vivo}. B) The selection probability of each layer for the optimal model from a comprehensive model space. Error bars represent the 95\% confidence interval from five independent search runs. Colors are matched to the layers in panel A.
    \revision{Abbreviations: conv\_k\{1,3,5\}, convolution with kernel size of 1, 3, 5; conv\_d4, convolution with kernel size=3 and dilation rate=4; conv\_lin, convolution with kernel size=1 and linear activation function. bidirection\_\{32,16\}: bidirectional RNN layer with 32 or 16 hidden units. kinn\_\{$ij$\}: searched KINNs with $i$ enzymatics states trained on the $j$-th \textit{in vitro}  gRNA dataset.}
  }
  \label{fig:elektrum_structure}
\end{figure}

\subsection{Novel integration of transfer learning with architecture search boosts \textit{in vivo} cleavage prediction}
\label{sub:transfer_learning}
We investigated if we could augment the KINN's ability to predict Cas9 cleavage kinetics \textit{in vivo} while maintaining its mechanistic interpretability. To better capture Cas9 cleavage kinetics \textit{in vivo}, we employed \textit{Elektrum}’s transfer learning strategy that incorporates a trained KINN with a convolutional backbone (Figure \ref{fig:elektrum_structure}). This convolutional backbone is built by a CNN architecture search algorithm \cite{Zhang2021AMBER} to account for sequence context effects that are not captured in the \textit{in vitro}-learned kinetic rates (Methods). Furthermore, multiple candidate KINNs are searched to automatically determine the optimal kinetic model that renders the most predictive power for the \textit{in vivo} datasets.
\revision{Thanks to the full interpretability of KINNs, the convolutional backbone only needs to learn a modifier term for \textit{in vivo}-specific effects while keeping the pretrained weights of \textit{in vitro} KINNs frozen.
  While fine-tuning KINN weights may improve the final model's performance, it can complicate the interpretation of the original KINN parameters and reduce interpretability. With that said, there is an argument to enable fine-tuning in KINN in our code implementation, making it easy to adapt for other use cases.}
Finally, we constructed a model ensemble by running the (stochastic) NAS multiple times (n=5). We applied this transfer learning strategy to enable the integration of the mechanistic \textit{in vitro} trained KINN models with the sequence context learned from large \textit{in vivo} data collections \cite{lin2020crispr}. Our transfer learning architecture search substantially improved the \textit{in vivo} performance of cleavage probability prediction as compared to the base KINN trained on \textit{in vitro} data alone. The \textit{Elektrum} model ensemble improves testing performance across the validation (Extended Data Figure \revision{5}) and all external datasets;
this approach even outperformed the current non-interpretable state-of-the-art predictors in 2 out of the 3 test datasets (Table \ref{tab:compare}).

Surprisingly, we found that, from a comprehensive space of possible CNN models, the data-driven NAS consistently selected (across all independent search runs) simpler architectures. The search algorithm was trained to select from variable kernel sizes for two convolution layers and included a specialized Inception layer \cite{szegedy2016rethinking}. This was followed by a selection between a simple and a complex bi-directional Long-Short Term Memory (LSTM) layer, and between a standard and a self-attention flatten layer (see Methods for details; Figure \ref{fig:elektrum_structure} and Extended Data Figure \revision{6}).
This broad set of architectures covers most previously proposed expert-designed CNNs \cite{Zhou2015, Li2021b, lin2020crispr, Kelley2018}.
\revision{The searched optimal architectures for each of the five runs are detailed in Extended Data Table 3.}
As a result, the simpler architectures selected by the data-driven NAS used the first convolution layer with smaller kernel sizes instead of the more sophisticated Inception layer chosen by human experts in ref. \cite{lin2020crispr}.  Furthermore, the second convolution layer, common in human-designed CNNs \cite{Zhou2015, Kelley2018}, tended to be replaced by an identity layer and thus removed by NAS. In contrast to simplifying convolution layers, the NAS found it important to select the complex over the simple bi-directional LSTM layer (Figure \ref{fig:elektrum_structure}), suggesting the potential existence of long-range interactions in this biological system.

\begin{figure}[h]
  \centering
  \includegraphics[width=0.9\textwidth]{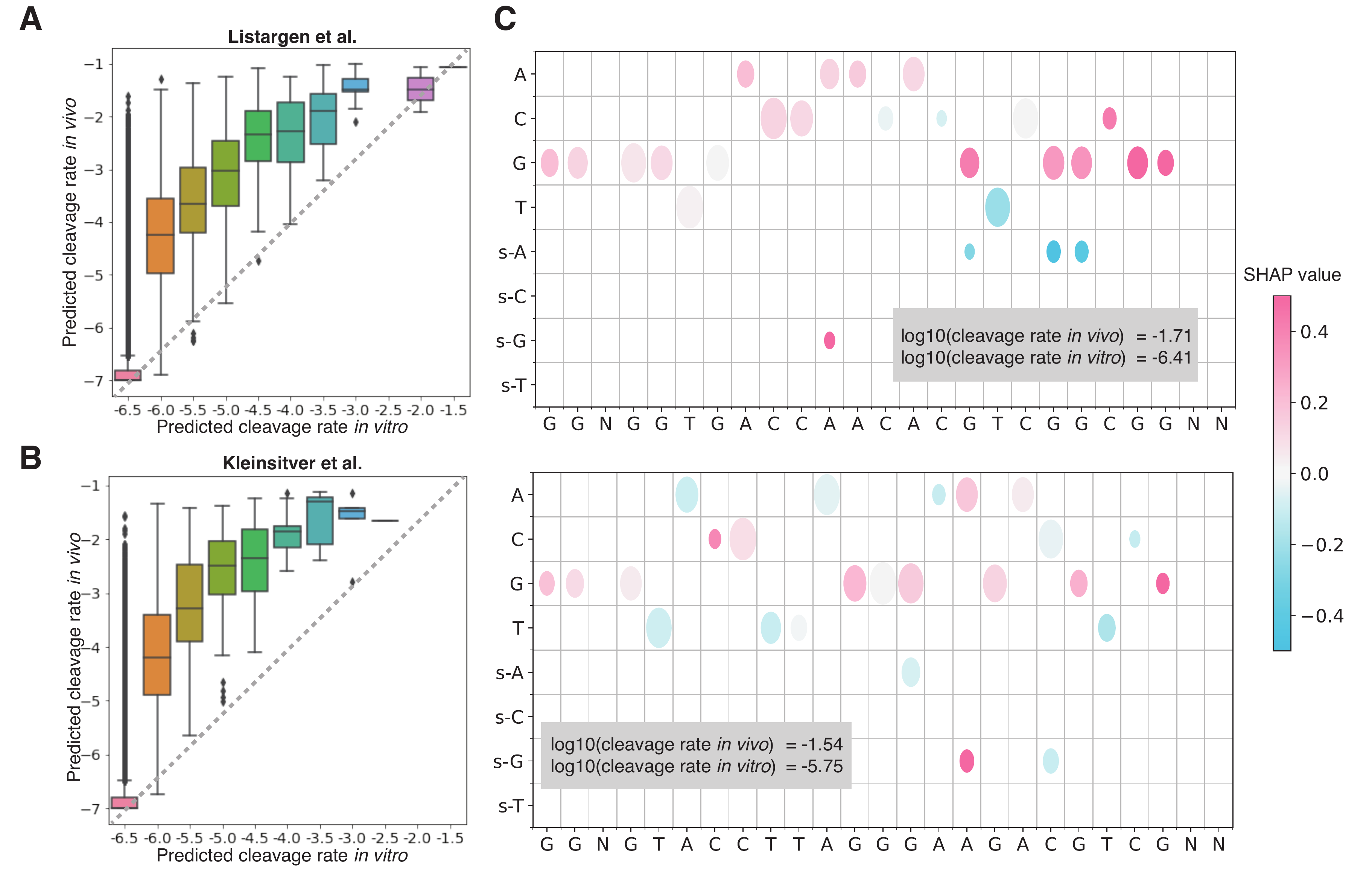}
  \caption{Interpreting sequence context for \textit{in vivo} Cas9 cleavage rate prediction. The comparison of the full \textit{Elektrum} \textit{in vivo} predicted cleavage rates versus the KINN \textit{in vitro} predicted cleavage rates in A) Listargen data \cite{listgarten2018prediction} and B) Kleisitiver data \cite{kleinstiver2015engineered}. Both datasets were held-out during the training of \textit{Elektrum} and KINN. Boxplot elements: center line, median; box limits, upper and lower quartiles; whiskers, 1.5x interquartile range; points, outliers. C) SHAP analysis reveals the presence of guanines in the Cas9 target regions is important to increased cleavage \textit{in vivo}. Shown are two gRNA-DNA pairs with the largest predicted cleavage rate difference between \textit{in vivo} and \textit{in vitro}. The gRNA sequence is shown on the x-axis. \revision{On the y-axis, the first four rows represent one-hot encoded gRNA sequence, and the last four rows represent the observation of nucleotide substitutions at specific locations in the target DNA sequence.} Color represents the average SHAP values across five best models created independently in five search runs; size of the dots scale inversely with variance of SHAP across the five models.}
  \label{fig:shap}
\end{figure}

\subsection{Interpreting importance of sequence context for \textit{in vivo} Cas9 cleavage rate prediction}
\label{sub:elektrum_interpretation}

A unique advantage of \textit{Elektrum}, besides superior performance, is its prediction of the physical \textit{in vivo} kinetic rates. This enables us to assess which sequence contexts are important for Cas9 \textit{in vivo} editing by comparing the full \textit{Elektrum} \textit{in vivo} model predictions versus the KINN \textit{in vitro} predictions (Figure \ref{fig:shap}).  We found a general strong positive correlation between the \textit{in vitro} cleavage rates and \textit{in vivo} predicted cleavage probabilities, suggesting that \textit{in vitro} kinetics is relevant to the \textit{in vivo} context.
\revision{However, despite the overall positive correlation between \textit{in vitro} and \textit{in vivo} rates, \textit{in vivo} rates were generally higher than \textit{in vivo} rates (blue, green and orange boxes in Figure \ref{fig:shap}A).
}
The underestimation for cleavage rate \textit{in vitro} was consistent across two external datasets (Figure \ref{fig:shap}A and \ref{fig:shap}B), suggesting the involvement of particular sequence elements in \textit{in vivo} cleavage activation.

To understand the sequence determinants underlying these differences between \textit{in vivo} vs \textit{in vitro} predicted Cas9 cleavage rate, we assessed the context-specific base-pair importance using SHAP \cite{NIPS2017_7062}. SHAP is an explainable AI technique based on game theory that quantifies each feature’s contribution to a model’s prediction (Methods). SHAP analysis here indicates that the presence of guanines in the Cas9 target regions is important to explain the increased cleavage seen \textit{in vivo} (Figure \ref{fig:shap}). Across all analyzed sequences, guanines had significantly higher SHAP values compared to other nucleotides in the target region (Extended Data Figure \revision{7}).
This suggests that guanine-rich off-target regions are more likely to be cleaved \textit{in vivo}, a phenomenon that is concordant with previous reports \cite{moreno2015crisprscan}.

\section{Discussion}
In the biological sciences, developing quantitative understanding and building predictive mechanistic models requires data from both purified systems and cellular data.  In purified \textit{in vitro} systems, most components are known and their interactions are simpler to understand, but the assays lack relevant cellular or genomic context. On the other hand, \textit{in vivo} cellular data contains this context and is often more plentiful, but that context and measurement uncertainties make it difficult to identify underlying mechanisms and confounding effects. Biophysical and kinetic models can elucidate underlying mechanisms, while deep learning models are powerful tools and need little prior knowledge for making predictions. By combining KINN ensemble generation with NAS-selected transfer learning, \textit{Elektrum} leverages both data sources to generate interpretable mechanistic models that also integrate genomic and cellular context.

The overall utility of this approach is demonstrated by \textit{Elektrum}’s performance compared to more traditional state-of-the-art NN models. \textit{Elektrum} is both interpretable and predictive because of the information encoded in the basic physics-informed structure of the KINNs. Furthermore, in allowing the NAS to choose from an ensemble of high-quality KINNs, we can assess and fine-tune over this ensemble of kinetic models in a data-driven manner using a transfer learning approach.

Our method was designed to rapidly hypothesize an ensemble of kinetic models and then apply said models with little prior knowledge to \textit{in vivo} processes. The interpretable NAS is efficient because KINNs are backwards differentiable and train faster than other parameter optimization methods \cite{Tareen2019, Tareen2021}.
However, because \textit{in vivo} systems are affected by biological factors not present in \textit{in vitro} systems, accurate predictions depend on the learned enzymatic reaction being a key step in the \textit{in vivo} biological process. For example, we assume the kinetics of DNA cleavage is the dominant process for predicting Cas9 editing \textit{in vivo}. If this assumption is violated, the model should require a larger CNN backbone to compensate for processes not encapsulated in the enzymatic reaction. However, \textit{Elektrum} tended toward lower complexity CNNs during NAS (Figure \ref{fig:elektrum_structure}), supporting the idea that the cleavage pathway learned by KINNs is paramount in \textit{in vivo} editing.

\revision{\textit{Elektrum} provides a powerful tool for prioritizing gRNA targets to enable safe and effective genome editing. Given a gRNA sequence, \textit{Elektrum} predicts the cleavage rates for both on-target and off-target regions. The advantage of \textit{Elektrum} is its state-of-the-art predictive performance as well as its full interpretability. However, \textit{Elektrum} still makes false positive and false negative predictions when using \textit{in vivo} experimentally measured off-targets as gold standards. We performed the SHAP analysis for the top false postive and false negative predictions of the \textit{in vivo} model (Extended Data Figure 8) and \textit{in vitro} model (Extended Data Figure 9). The false negatives (experimentally measured off-targets but predicted unedited) have multiple mismatches, sometimes even on the PAM sites, suggesting additional factors that may have facilitated the \textit{in vivo} cleavage of these loci.}

\revision{The interpretability of the KINN models lends biophysical insights into possible contributing factors to inaccurate predictions. For example, in Moreb and Lynch \cite{moreb2021genome}, the authors argue that the lack of model generalization between species is caused by the variability of Cas9/gRNA "search spaces", the locations and states enzymes must explore to find correct target sequences. Equipped with sequence dependent kinetics, the search space can be defined as an organism's genome modeled as a bath of various DNA substrates. As will be presented in future work, off-target sites with large Cas9/gRNA dwell times sequester the protein complexes from solution, slowing down the cleavage rate of target sites; a phenomenon seen previously but not fully characterized from experimental work \cite{moreb2020crispr}. Furthermore, unlike conventional machine learning models, these KINN-informed biophysical models can easily include known processes like protein or RNA degradation, further changing Cas9 cleavage efficiency in an off-target concentration dependent manner. Such considerations help unify the discrepancies between species specific predictions. Still, more studies are required to elucidate other important \textit{in vivo}-specific effects, such as epigenetic modification, accessibility of genomic regions, and cell line-specific effects.}

\textit{Elektrum}’s method for discovering reaction kinetics only requires sequence-like input (amino acid sequence, DNA sequence, epigenetic marks, etc.) that can be perturbed and leads to different observable and time-dependent outcomes. Massively parallel kinetic profiling assays \cite{Jones2021, shen2021kinetic}
are prime candidates for \textit{in vitro} data for training KINNs. As more MPKPs are published, the \textit{Elektrum} framework will become a common tool to quickly discover kinetic models and predict related biological pathways. In the future, \textit{Elektrum} could be extended to other protein and irreversible biochemical reactions such as the kinetics of restriction enzymes, proteases, motor protein ATP cycles and motion, and ribosomes.

\section{Methods}\label{sec11}

\subsection{Kinetic theory and King-Altman diagrams}\label{sec:kinetic_theory}

Modeling enzymatic reactions gives us a way to predict the production, degradation, and transition of biochemical components into different states. Here we first describe the theoretical foundations for chemical kinetic theory and then translate governing equations to a neural network optimization problem.

A linear multi-step enzymatic reaction can be defined using a system of equations in the form of
\begin{equation}
  \label{eq:dS}
  \frac{\partial S_\alpha}{\partial t} = \sum_{\beta}k_{\beta\alpha}S_\beta - S_\alpha \sum_{\beta} k_{\alpha\beta},
\end{equation}
where $k_{\alpha\beta}$ is the transition rate constant from enzymatic state $\alpha$ to $\beta$ and $S_\alpha$ is the concentration of reactant or product in state $\alpha$. The values of reaction rates depend on a discrete sequence vector $\bx$.
Equation \eqref{eq:dS} can be represented as a graph $\mathcal{G} = (V,E)$ where vertices $V \subseteq \{S_\alpha\}$ and edges $E \subseteq \{k\ab | S_\alpha,S_\beta \in V\}$. All reactions of this kind then exist in the space $\mathbb{G} = \{\mathcal{G}\}$. Our goal is to search $\mathbb{G}$ to determine the correct $\mathcal{G}$ for a reaction given that the system is in a pseudo-steady state and that there are measurable products from specific reactions $\tilde{k}\ab$, where $\tilde{k}\ab$ is a subset of reactions that lead to an experimentally measurable quantity.

Equation (\ref{eq:dS}) can be written more succinctly by defining the matrix
\begin{equation}
  \label{eq:bK}
  (\bK)\ab = k_{\beta\alpha} - \delta\ab\sum_\gamma k_{\alpha\gamma}
\end{equation}
and, for convenience, normalizing the concentration vector
\begin{equation}
  \label{eq:bs}
  (\bs)_\alpha = \frac{S_\alpha}{\sum_\beta S_\beta}
\end{equation}
so that
\begin{equation}
  \label{eq:dbs}
  \pder[\bs]{t} = \bK\cdot\bs,
\end{equation}
which has the solution
\begin{equation}
  \label{eq:Ssol}
  \bs(t) = \bs_{\infty} + \left( \bs(0) -\bs_{\infty}\right) e^{-\bK t}
\end{equation}
where $\bs_{\infty}$ is the normalized steady-state solution, i.e., $\bK\cdot\bs_{\infty}=0$.  We can solve for components of $\bs_{\infty}$ using the King-Altman (KA) method \cite{King1956}.

The King-Altman method uses Cramer's rule to show that
\begin{equation}
  \label{eq:salpha}
  (\bs_{\infty})_\alpha = \frac{\sum_{\ell \to \alpha}\kappa_\ell}{\sum_{\ell'}\kappa_{\ell'}},
\end{equation}
where
\begin{equation}
  \label{eq:ka}
  \kappa_\ell = \prod_{ij \in \ell} k_{ij}= \exp \left[ \sum_{ij \in \ell} \log(k_{ij}) \right],
\end{equation}
where the product is over all the rates in the $\ell$th KA diagram. KA diagrams are acyclic subgraphs of $\mathcal{G}$ that contain edges one less than the number of all vertices in $\mathcal{G} $\cite{CornishBowden1977}. Also, all directed edges (rate constants) must be a part of a path that leads to the same vertex (state) in that diagram. The notation $\ell \to \alpha$ means that KA diagram $\ell$ ends on state $\alpha$. Given an arbitrary $\bK$, we find all KA diagrams using the algorithm put forth by Lam and Priest \cite{Lam1972}.

There exists a subset of reactions with rate constants $\tilde{k}\ab \subset k\ab$ that map to an experimentally measurable quantity $y$. The sum of all the production rates is called the `activity' denoted as $\act$:
\begin{equation}
  \label{eq:act}
  \act \propto \sum_{\alpha\beta \in {\rm act}} \tilde{k}\ab s_\alpha,
\end{equation}
where only certain state changes, i.e., $\alpha\to\beta \in {\rm act}$, contribute to activity and map to an experimental measurable with the equation $y = \sigma(\act)$, where $\sigma(x)$ is a one-dimensional monotonically increasing function with domain and range $\in [0, \infty)$. The model search space includes all possible $\tilde{k}\ab$ and $\sigma$ so that $\mathcal{G} \in \left\{\mathbb{G}, \{\tilde{k}\ab\}, \{\sigma\}\right\}$.

\subsection{Construction of kinetically interpretable neural networks from the King-Altman formalism}
\label{sec:construct_KINNs_with_KA}
Kinetically Interpretable Neural Networks (KINNs) are constructed as a more generalized framework based on the previous work \cite{Tareen2019}.
Given the number of enzymatic states, we focus on optimizing KINN's hyperparameters that determine the mapping of sequence to kinetic rate constants. We perform a separate analysis for each specified number of enzymatic states and associated transition rates, with the number of states ranging from 4 to 6. Automatic exploration of viable multistate models is left for another paper.  A schematic of a KINN architecture and the associated 4-state kinetic model for Cas9-DNA cleavage is shown in Extended Data Figure \revision{10}.

The first KINN layer aims to learn the mapping of sequence $\bx$ to kinetic rate constants $k_{\alpha\beta}$, $\forall \{\alpha,\beta\}$.
We note that kinetic rate constants are specific variables describing a reaction independent of reagent concentrations (as opposed to time-dependent kinetic rate), although the term `constant' may cause potential confusions for readers unfamiliar with this concept.
We assume a rate constant $k_{\alpha\beta}$ is dependent on a subsequence of $\bx$, denoted by $x_{i:j} \in \mathbb{R}^{(j-i)\times4}$, the one-hot encoded subsequence from the $i$-th to the $j$-th nucleotide on the input sequence $\bx_{i:j}$.
We use a single-filter convolutional layer, with kernel $W_{\alpha\beta} \in \mathbb{R}^{d \times 4}$
and bias term $b_{\alpha \beta}$, to learn $k_{\alpha\beta}$:
\begin{equation}
  \log(k_{\alpha\beta})=\sum_i^j Conv(x_{i:j}; W_{\alpha\beta}) + b_{\alpha\beta},
\end{equation}

For each $k_{\alpha\beta}$, we search the hyperparameters including the sequence determinant index $i,j$ and the convolution kernel size $d$. When $d=1$, the operation simply computes a weighted sum of each gRNA-DNA nucleotide pair's contribution. When $d>1$, it can model the local neighbor interaction effects within the window. When $d=(j-i)$, the kernel length is equal to the input sequence length, and the convolution layer becomes a fully connected layer.

The second KINN layer nodes correspond to the KA diagram terms $\kappa_{\ell}$ in equation \eqref{eq:ka}. The first and second layers are connected by a static(untrained) binary layer $\bB$ with the connectivity determined by the transitions in each KA diagram described previously \cite{Lam1972}.
An example of KA diagrams and associated binary matrix for a 3-state cas9 system is shown in Extended Data Figure \revision{11}.
The summation of rates in log-space implies a multiplication in linear-space. Therefore, equation \eqref{eq:salpha} can be re-written by applying a softmax activation to the product of the rate constant layer with $\bB$
\begin{equation}
  \bs_\infty = \mathrm{softmax}(\bB \cdot \log(k\ab)).
\end{equation}

Finally, the effective cleavage rate is computed from equation \eqref{eq:act}. In our Cas9 massively parallel kinetic profiling assay, this is the observed cleavage rate. To calculate activity, exponentiated log-rate nodes' values from the first layer multiply the enzymatic state before the activity-producing transition. If multiple rates contribute to activity, then all activity rates are included according to equation \eqref{eq:act}. A final activation function or mapping may be applied to the activity node if a measurement of the activity-measurement relation is known to be non-linear.
\begin{equation}
  \label{eq:y}
  y = \sigma(\act) = \sigma_o ( \sigma_\act(\act \cdot W_t + b_t) \cdot W_o + b_o ),
\end{equation}
where the non-linear activation functions $\sigma_\act$, $\sigma_o$, and dimensions of $W_t$, can also be searched.


\subsection{Probabilistic model building genetic algorithm for searching KINN}
\label{sec:prob_model_search}
We employ a genetic algorithm to search for KINN model architectures. To incorporate existing knowledge and prior beliefs about a kinetic system, we use a Bayesian genetic algorithm based on probabilistic model building \cite{pelikan2011probabilistic}. For each kinetic rate constant, we first define a prior probability distribution for its hyperparameters: the sequence determinant $i,j$ and the convolution kernel size $d$. Then we sample model architectures and train the weights to evaluate performance. The search algorithm updates the posterior distribution by the architectures of each generation's surviving models.

Following the notations of $k_{\alpha\beta}$ and its sequence determinants $x_{i:j}$, as an uninformative prior, we set $i \sim$ Multinomial($[\phi(k)-w/2, \phi(k)+w/2]$) and $(j-i) \sim$  Multinomial([1,5]), where $\phi(k$) is the anchor function that uniformly maps each kinetic rate to sequence intervals on the input sequence $\bx$; $w$ is the search window size. For instance, for a four-state {$s_0, s_1, s_2, s_3$} KINN with 20bp input sequence $x$, $\phi(k_{0 1})=5$, $\phi(k_{1 2})=10$, and $\phi(k_{2 3})=15$. The function that maps $x_{i:j}$ to $\log(k_{\alpha\beta})$ is parameterized as a simple one-layer linear convolutional neural network and we search for the convolution kernel sizes.

  In the first iteration, the genetic algorithm uniformly samples a number of $m=10$ KINNs from the prior probability distributions. The fitness of a KINN is defined by Pearson correlation between the prediction and observations from the validation dataset. The expected fitness for each generation $t$, $C_t$, is evaluated by the average fitness values of KINNs sampled. Subsequently, KINNs with a fitness higher than $C_t$ survive and their model architectures are used to update the posterior distributions.

  \subsection{Benchmarking by simulated kinetic datasets}
  \label{sec:benchmarking}
  We tested our method of KINN optimization on simulated data where the kinetic rates' dependence gRNA-DNA alignment was known exactly. We started with an arbitrary sequence of 50 nucleotides. Only a subsequence from the 5th to the 40th nucleotide pairs contribute to the simulated cleavage rate. The 35 nucleotide pairings were separated into seven five-nucleotide long subsequences, each determining the value of a different kinetic rate constant for a 4-state Cas9 cleavage cycle.

  Transition rates were calculated from the Arrhenius equation
  \begin{equation}
    \label{eq:forward_k}
    k\ab = k_o \exp \left[ \mp (k_bT)^{-1} \sum_{i\in\Omega\ab} \Delta \bar{G}\ab(x_i, \xi_i) \right]
  \end{equation}
  where the factor inside the exponent is negative for forward reactions ($\alpha < \beta$) and positive for backwards reactions ($\alpha > \beta$). $\Omega\ab$ is the set of pair-indices $i$ that contribute to a reaction’s free energy barrier. No rates shared nucleotide pair dependence with other rates, i.e., $\Omega\ab \cap \Omega_{\alpha'\beta'}=\emptyset $ for $\alpha, \beta \neq \alpha', \beta'$.  $\Delta \bar{G}\ab(x_i, \xi_i)$ is the difference in free energy contributed by the $i$th DNA-gRNA nucleotide pair for the transition $\alpha \to \beta$. For this data, the values were taken from the set $\Delta \bar{G}\ab(x_i, \xi_i) \in \{-1, -0.1, 0.1, 1\}$. Complementary pairs were assigned a value of $-1$ but mismatched pairs varied for different transitions. All rates were multiplied by a baseline kinetic rate $k_o=.01$ sec$^{-1}$. Rate constant $k_{01}$ is technically multiplied by $k_o=.01$ sec$^{-1}$nMol$^-1$ to maintain proper units.

  Once the rate dependence was specified, 20000 sequences were randomly generated and the forward and backward rates calculated. From these rates, we constructed the rate matrix from equation (\ref{eq:bK})
  \begin{equation}
    \label{eq:Kcas9}
    \bK_{cas9} = \left(
    \begin{array}{cccc}
        -k_{01}D_u & k_{10}           & 0                & k_{30}           \\
        k_{01}D_u  & -(k_{10}+k_{12}) & k_{21}           & 0                \\
        0          & k_{12}           & -(k_{21}+k_{23}) & k_{32}           \\
        0          & 0                & k_{23}           & -(k_{30}+k_{32}) \\
      \end{array}
    \right),
  \end{equation}
  where $D_u=1$ nMol is the constant concentration of unbound DNA. The largest, non-zero eigenvalue was deemed the cleavage rate for DNA-gRNA pairs, i.e., what KINNs were trained to predict.

  \subsection{Relation between King-Altman and single rate learning}%
  \label{sub:relation_between_king_altman_and_single_rate_learning}
  We observed that training our KA network on single-rate data produces
  accurate predictions of the kinetic rate constants and activity despite not being at steady state, a requirement for the KA method.  This suggests some equivalency between the cleavage rate observed by steady state and uncut DNA-depleting systems. Here we show that, under certain assumptions, the data produced by the two systems are approximately equivalent for training KINNs.

  First we construct a kinetic model with a depleting uncut DNA reservoir $D_u(t)$
  \begin{equation}
    \label{eq:K'}
    \bK' = \left(
    \begin{array}{cccc}
        -k_{01}S_0 & k_{10}           & 0                & 0                \\
        k_{01}S_0  & -(k_{10}+k_{12}) & k_{21}           & 0                \\
        0          & k_{12}           & -(k_{21}+k_{23}) & k_{32}           \\
        0          & 0                & k_{23}           & -(k_{34}+k_{32}) \\
      \end{array}
    \right)
  \end{equation}
  with unnormalized state vector $\bS(t) = (D_u,S_1, S_2, S_3)^T$. Notice the differences between the matrices (\ref{eq:Kcas9}) and (\ref{eq:K'}). The element $(\bK')_{3,3}=-(k_{34}+k_{32})$ while $(\bK_{cas9})_{3,3}=-(k_{30}+k_{32})$, there are elements of $k_{01}S_0$ in $\bK'$ instead of $k_{01}D_u$ as in $\bK_{cas9}$, and finally $(\bK')_{0,3} = 0$ while $(\bK_{cas9})_{0,3} = k_{30}$. The first difference is just notational since both rates produce the measurable quantity, cut DNA, and are therefore equivalent $k_{34} = k_{30} = k_{\rm cut}$. A similar argument can be made for the second difference assuming that the concentration of unbound Cas9 $S_0$ remains roughly constant. This allows us to swap $D_u$ for $S_0$. However, the final difference changes the determinant of the matrix
  $$\det{\bK'}\neq0 \quad \text{while} \quad \det{\bK_{cas9}}=0.$$
  If we compare the characteristic polynomials $C(x;\bK)$ of these models
  $$C(x;\bK) = \sum_{n=0}^{N} a_n(\bK)x^n = \prod_{n=1}^N(x - \lambda_n),$$
  we see that $C(x;\bK')-C(x;\bK_{cas9}) = a_0(\bK') = \det{\bK'}$. This is notable because $\det{\bK'} = k_{01}k_{12}k_{23}\kcut = \sum_{\alpha\beta \in {\rm act}} \tilde{k}\ab s_\alpha$ of the original system. Even more interesting is that $a_1$ for both systems is the denominator of equation (\ref{eq:salpha}),  i.e.
  $$a_1(\bK) = a_1(\bK') = \sum_{\ell'}\kappa_{\ell'}.$$
  Therefore, small $|\lambda|$ values are approximately
  \begin{equation}
    \label{eq:lambda_approx}
    0 = C(\lambda;\bK') = a_0 + a_1\lambda + O(\lambda^2)
  \end{equation}
  so
  \begin{equation}
    \label{eq:eig_ka_equiv}
    - \lambda \approx \frac{a_0}{a_1} = \frac{\sum_{\alpha\beta \in {\rm act}}\tilde{k}\ab s_\alpha}{\sum_{\ell'}\kappa_{\ell'}},
  \end{equation}
  which is the activity $\act$ given by the King-Altman equations (\ref{eq:salpha}) and (\ref{eq:act}). However, for more complicated systems, this approximation may not hold. Therefore, one would utilize an eigendecomposition layer instead of the KA layer to find the largest eigenvalue\cite{Wang2019}. This approach was tested and showed similar performance to KA layers (Extended Data Figure 1).

  \subsection{Convolutional neural network architecture search with pre-trained KINNs}
  \label{sec:conv_pretrain_kinns}

  We developed a new neural architecture search (NAS) method for automatically searching CNNs with pre-trained KINNs to model \textit{in vivo} kinetic systems. Let $\bm{\kappa}=\log(\bk)$ be a vector of log kinetic rate constants from a pretrained KINN \textit{in vivo}, and $\bk$ be the kinetic rates \textit{in vitro}, we seek to learn a modifier term $\delta_\ell$ for each kinetic rate $k_\ell$ such that $\delta_\ell$ account for the \textit{in vivo} sequence-context effects and $\bk' = \exp(\bm{\kappa}+ \boldsymbol{\delta})$.
  These \textit{in vivo} kinetic rates $\bk'$ are then fed into the downstream KINN layers of King-Altman layer and activity layer. The final binary output of cleavage probability is modeled by a sigmoid layer with a single scalar kinetic activity as input.

  We parameterized the functional dependency of $\boldsymbol{\delta}$ to the input gRNA and DNA sequences as a deep convolutional neural network backbone $f$.
  The input for $f$ is a 13-bit encoding for gRNA-DNA alignment by 25 nucleotides, where we used 4-bit to one-hot encode the gRNA sequence, 4-bit to one-hot encode DNA substitutions, 4-bit to one-hot encode DNA insertions, and 1-bit to encode DNA deletions.
  The 25 nucleotides consist of 3bp PAM site, 20bp gRNAs, and 2bp for padding DNA insertions.
  \revision{The padding columns are represented as Ns and are encoded as zeros.}
  The output from the CNN is a flatten vector, which will be linearly transformed to $\boldsymbol{\delta}$ and followed by $\log(\bk') = \boldsymbol{\kappa} + \boldsymbol{\delta}$.
  The dimensionality of the linear transformation will be determined based on the pretrained KINN and its $\bm{\kappa}$.
  The pretrained KINNs are converted to special layers with two inputs: one input directly from the gRNA-DNA alignments to compute $\bm{\kappa}$, and the other input $\boldsymbol{\delta}$ from the deep CNN backbone.

  To search for optimal CNN architecture, we employed AMBER (v0.1.3), an AutoML framework developed previously \cite{Zhang2021AMBER} that demonstrated state-of-the-art NAS performance in genomics. AMBER searches neural network architectures by two components, 1) a controller model that optimizes neural architecture through reinforcement learning and 2) a model space to sample neural architectures. The search and optimization process of neural network architectures is described in length previously \cite{Zhang2021AMBER}.

  Our model space for the CNN architectures consisted of 7 layers. All layer search spaces besides the first convolution layer, the flatten layer, or the KINN layer, include an identity operator which would effectively remove that layer if selected. This enabled the controller model to reduce model complexity during architecture optimization. The first two layers are convolution operators with different kernel sizes and dilation rates. Alternatively,  the controller can select specialized Inception layers that aim to capture patterns at multiple scales by combining kernels of different sizes into a single layer. The convolution layer was followed by a dropout layer to regularize model complexity and improve model generalization, where we searched for its dropout probabilities. The next layer searched for bidirectional LSTM layers with either 32 units or 16 units, with the option of an identity operator. This layer was followed layer with a search space that included a standard flatten layer, an attention-based flatten layer, a dense layer of 64 units, and an identity operator. Finally, we included the KINNs pretrained on the \textit{in vitro} data, with three different state numbers on two separate datasets, yielding a total of 6 pretrained KINN configurations.

  \subsection{CRISPR/Cas9 datasets and train-testing split}
  \label{sec:train-test_split}

  To create the \elektrum~model, we divided the CRISPR/Cas9 cleavage datasets into two categories. The first category consisted of time-resolved kinetic assays measured \textit{in vitro}. The second category consisted of \textit{in vivo} off-target editing measurement using high-throughput assays. We described the processing and train-test split for each category below.

  For the \textit{in vitro} data, we re-compiled the Massively parallel kinetic profiling (MPKP) data for Cas9 generated previously \cite{Jones2021}. The MPKP measured the cleavage rate of SpCas9 on ~14,000 targets containing mismatches, insertions, and deletions relative to two different gRNA templates.  For each gRNA-DNA pair, we converted their sequences as described in the previous section to as a 13-bit by 25-bp encoding matrix. For each gRNA sequence, we trained separate models using all data points, while using 50\% of the other sgRNA datapoints as validation data in order to search for neural network architectures. The remaining 50\% of the other sgRNA data points are held out as testing data.

  For the \textit{in vivo} data, we used the assembled training dataset from CRISPR-Net \cite{lin2020crispr}. Briefly, this training data was collected and uniformly processed from 8 datasets in previously published Cas9 off-target assays \cite{tsai2017circle, listgarten2018prediction, doench2016optimized, haeussler2016evaluation, cameron2017mapping, kleinstiver2015engineered}. In this data, the negative data points (that is, inactive off-target sites) were constructed by searching genome-wide DNAs by Cas-Offinder \cite{bae2014cas}, a computational tool that searches for potential off-target sites of Cas9. To ensure a fair comparison, we held out the same three testing datasets to match the original train-test split as in CRISPR-Net. In total, the testing dataset has 693,235 gRNA-DNA pairs. For the remaining data, we randomly split the dataset by gRNA sequences, holding out 20\% of the gRNA and all their corresponding gRNA-DNA pairs as validation datasets. This corresponds to a total of 842,298 gRNA-DNA pairs for training and 269,900 gRNA-DNA pairs for validation.

  \subsection{Neural architecture analysis for KINN and \elektrum}
  \label{sec:arch_anal_elektrum}

  To evaluate the neural architectures created from the KINN search on the Cas9 MPKP assays and account for the stochasticity between different search runs, we developed a Baysian hierarchical model to infer the maximum \textit{a posteriori} estimate for neural architecture parameters. The Bayesian hierarchical model allowed us to consider the uncertainty and covariance in a principled framework. Specifically, we employed a multivariate normal (MVN) distribution to model the group-level KINN architecture parameters. The MVN's mean followed an uninformative prior. Its variance-covariance matrix was defined by a distribution over Cholesky decomposed covariance matrices, such that the underlying correlation matrices followed an LKJ distribution \cite{lewandowski2009generating} with a standard deviation $\eta=1$. We assumed each replicate run draws a set of architecture parameters from the group-level MVN distribution, such that the optimized KINNs follows another sample-level MVN with its mean determined by the group-level sampled parameters. The covariance matrix of the sample-level MVN also followed an uninformative prior of an LKJ distribution with the standard deviations $\eta=1$. The top 5\% performing architectures from each run were fed into the Bayesian model for parameter fitting. We implemented the above model using PyMC3 \cite{Salvatier2016}.

For the transfer learning NAS selection probability, we simply pooled the last five architectures from each search run across five independent search runs. This was done because each layer was modeled by a recurrent neural network from its preceding layers without complex multivariate dependencies. We then computed the mean and variance of the selection likelihood for a candidate layer within the given model space.

\subsection{Base-pair importance interpretation by SHAP}
\label{sec:shap_interp}

In order to determine the base-pairs that are important in explaining the difference between \textit{in vitro} and \textit{in vivo} cleavage predictions, we applied SHAP analysis to both the base KINN and the trained \textit{Elektrum} model. Both KINN and \textit{Elektrum} were end-to-end differentiable, which allowed us to use the GradientExplainer to assess the importance of each base-pair within each model. GradientExplainer seeks to interpret a model using expected gradients, which is an extension of integrated gradients \cite{NIPS2017_7062}. We then normalized the maximum absolute base-pair importance to 1, making the importance scores comparable across different predictions. After obtaining the normalized base-pair importance scores for both \textit{in vitro} and \textit{in vivo} cleavage predictions, we subtracted the \textit{in vitro} scores from the \textit{in vivo} scores to determine the specific importance of each base-pair in explaining the difference between \textit{in vitro} and \textit{in vivo} cleavage. In this final base-pair importance matrix, if a base-pair contribution is identical \textit{in vivo} vs \textit{in vitro}, it’s score would be zero; in contrast, a positive importance score indicates the activation of cleavage \textit{in vivo} specifically, and a negative value indicates the inactivation \textit{in vivo} specifically. To compute the integrated gradient values, we used an all-zero matrix of shape 25 by 13 as the background. Finally, using the five models constructed by independent AMBER runs, our interpretation method of base-pair effects considers both the averaged SHAP values as well as its variance across models.

\section{Data Availability}
\label{sec:data_avail}
All training and testing data presented in this paper are publicly available. The \textit{in vitro} kinetic data was downloaded from Jones et al. \cite{Jones2021} in URL \url{https://github.com/finkelsteinlab/nucleaseq/tree/master}. The \textit{in vivo} CIRSPR/Cas9 off-targeting dataset was downloaded from CRISPR-Net \cite{lin2020crispr} in URL \url{https://codeocean.com/capsule/9553651/tree/v2}.

\section{Code Availability}
\label{sec:code_avail}
The \textit{Elektrum} code is available on GitHub at \url{https://github.com/zj-zhang/Elektrum}.

\bmhead{Acknowledgments}
We thank Chandra Theesfeld and Erik Thiede for their useful conversations.

\section*{Declarations}

\begin{itemize}
  \item Funding: OT is supported by National Institutes of Health (NIH) grant R01GM071966, and Simons Foundation grant no. 395506.
  \item Conflict of interest/Competing interests: None
  \item Ethics approval: Not applicable
  \item Consent to participate: Not applicable
  \item Availability of data and materials: See above
  \item Code availability: See above
  \item Authors' contributions: ZZ and ARL conceived the study, wrote the code, and performed the analysis. MS and OT supervised the study. ZZ, ARL, MS and OT wrote the paper.

\end{itemize}









\bibliography{elektrum, added}

\end{document}


\title[Article Title]{Supplemental Material}

\author*[1]{\fnm{Zijun} \sur{Zhang}}\email{Zijun.Zhang@cshs.org}
\equalcont{These authors contributed equally to this work.}

\author*[2]{\fnm{Adam R} \sur{Lamson}}\email{alamson@flatironinstitute.org}
\equalcont{These authors contributed equally to this work.}

\author[2,3]{\fnm{Michael} \sur{Shelley}}\email{}

\author[2,4]{\fnm{Olga} \sur{Troyanskaya}}\email{ogt@genomics.princeton.edu}

\affil[1]{\orgdiv{Division of Artificial Intelligence in Medicine }, \orgname{Cedars-Sinai Medical Center}, \orgaddress{\street{116 N. Robertson Blvd}, \city{Los Angeles}, \postcode{90048}, \state{CA}, \country{USA}}}

\affil[2]{\orgdiv{Center for Computational Biology}, \orgname{Flatiron Institute}, \orgaddress{\street{162 5th Ave}, \city{New York City}, \postcode{10010}, \state{NY}, \country{USA}}}

\affil[3]{\orgdiv{Courant Institute of Mathematical Sciences}, \orgname{New York University}, \orgaddress{\street{251 Mercer Street}, \city{New York City}, \postcode{10012}, \state{NY}, \country{USA}}}

\affil[4]{\orgdiv{Lewis Sigler Institute for Integrative Genomics}, \orgname{Princeton University}, \orgaddress{\street{Carl Icahn Laboratory
South Drive}, \city{Princeton}, \postcode{08544}, \state{NJ}, \country{USA}}}



\maketitle

\section{Kinetic theory and King-Altman diagrams}\label{sec:kinetic_theory}

Modeling enzymatic reactions gives us a way to predict the production, degradation, and transition of biochemical components into different states. Here we first describe the theoretical foundations for chemical kinetic theory and then translate governing equations to a neural network optimization problem. 

A linear multi-step enzymatic reaction can be defined using a system of equation in the form of
\begin{equation}
  \label{eq:dS}
   \frac{\partial S_\alpha}{\partial t} = \sum_{\beta}k_{\beta\alpha}S_\beta - S_\alpha \sum_{\beta} k_{\alpha\beta},
\end{equation} 
where $k_\ab$ is the transition rate from enzymatic state $\alpha$ to $\beta$ and $S_\alpha$ is the concentration of reactant or product in state $\alpha$. The values of reaction rates depend on a discrete sequence $\bx$.  
Equation \eqref{eq:dS} can be represented as a graphs so that a specific reaction has the graph $\mathcal{G} = (V,E)$ where vertices $V \subseteq \{S_\alpha\}$ and edges $E \subseteq \{k\ab | S_\alpha,S_\beta \in V\}$. All reactions of this kind then exist in the space $\mathbb{G} = \{\mathcal{G}\}$. Our goal is to search $\mathbb{G}$ to determine the correct $\mathcal{G}$ for a reaction given that the system is in a pseudo-steady state and that there are measurable products from specific reactions $\tilde{k}\ab$, where $\tilde{k}\ab$ is a subset of reactions that lead to experimentally measured quantity.

Equation (\ref{eq:dS}) can be written more succinctly by defining the matrix
\begin{equation}
  \label{eq:bK}
  (\bK)\ab = k_{\beta\alpha} - \delta\ab\sum_\gamma k_{\alpha\gamma}
\end{equation}
and 
\begin{equation}
  \label{eq:bs}
  (\bs)_\alpha = \frac{S_\alpha}{\sum_\beta S_\beta}
\end{equation}
so that
\begin{equation}
  \label{eq:dbs}
  \pder[\bs]{t} = \bK\cdot\bs, 
\end{equation}
which has the solution
\begin{equation}
  \label{eq:Ssol}
  \bs(t) = \bs_{\infty} + \left( \bs(0) -\bs_{\infty}\right) e^{-\bK t}
\end{equation}
where $\bs_{\infty}$ is the steady-state solution, i.e., $\bK\cdot\bs_{\infty}=0$.  We can solve for components of $\bs_{\infty}$ using the King-Altman (KA) method.

The King-Altman method uses Cramer's rule to show that 
\begin{equation}
  \label{eq:salpha}
  (\bs_{\infty})_\alpha = \frac{\sum_{\ell \to \alpha}\kappa_\ell}{\sum_{\ell'}\kappa_{\ell'}},
\end{equation}
where
\begin{equation}
    \label{eq:ka}
    \kappa_\ell = \prod_{ij \in \ell} k_{ij}= \exp \left[ \sum_{ij \in \ell} \log(k_{ij}) \right]
\end{equation}
where the product is over all the rates in the $\ell$th KA diagram. KA diagrams are acyclic subgraphs of $\mathcal{G}$ that contain edges one less than the number of all vertices in $\mathcal{G}$(cite). Also, all directed edges (rate constants) must be a part of a path that leads to the same vertex (state) in that diagram. The notation $\ell \to \alpha$ means that KA diagram $\ell$ ends on state $\alpha$. Given an arbitrary $\bK$, we find all KA diagrams using the algorithm put forth by Lam and Priest \cite{Lam1972}.

There exists a subset of reactions with rate constants $\tilde{k}\ab \subset k\ab$ that map to an experimentally measurable quantity $y$. The sum of all the production rates is called the `activity' denoted as $\act$: 
\begin{equation}
  \label{eq:act}
  \act \propto \sum_{\alpha\beta \in {\rm act}} \tilde{k}\ab s_\alpha,
\end{equation}
where only certain state changes, i.e., $\alpha\to\beta \in {\rm act}$, contribute to activity and maps to an experimental measurable with the equation $y = \sigma(\act)$, where $\sigma(x)$ is a one-dimensional monotonically increasing function with domain and range $\in [0, \infty)$. The model search space includes all possible $\tilde{k}\ab$ and $\sigma$ so that $\mathcal{G} \in \left\{\mathbb{G}, \{\tilde{k}\ab\}, \{\sigma\}\right\}$.

\section{Construction of kinetic-interpretable neural networks from the King-Altman formalism}
\label{sec:construct_KINNs_with_KA}
Kinetic-interpretable neural networks (KINNs) are constructed as a more generalized framework based on the previous work \cite{Tareen2019}.
Given the number of enzymatic states, we focus on optimizing KINN's hyperparameters that determine the mapping of sequence to kinetic rates. We perform a separate analysis for each specified number of enzymatic states and associated transition rates, with the number of states ranging from 4 to 6. Automatic exploration of viable multistate models is left for another paper.  

The first KINN layer aims to learn the mapping of sequence $\bx$ to kinetic rate constants $k_{\alpha\beta}$, $\forall \{\alpha,\beta\}$. 
We assume the rate constant $k_{\alpha\beta}$ is dependent on a subsequence of $\bx$, denoted by $x_{i:j} \in \mathbb{R}^{(j-i)\times4}$, the one-hot encoded subsequence from the $i$-th to the $j$-th nucleotide on the input sequence $\bx_{i:j}$.
We use a single-filter convolutional layer, with kernel $W_{\alpha\beta} \in \mathbb{R}^{d \times 4}$ 
and bias term $b_{\alpha \beta}$, to learn $k_{\alpha\beta}$:
$$\log(k_{\alpha\beta})=\sum_i^j Conv(x_{i:j}; W_{\alpha\beta}) + b_{\alpha\beta}$$

For each $k_{\alpha\beta}$, we search the hyperparameters including the sequence determinant index $i,j$ and the convolution kernel size $d$. When $d=1$, the operation simply computes a weighted sum of each gRNA-DNA nucleotide pair's contribution. When $d>1$, it can model the local neighbor interaction effects within the window. When $d=(j-i)$, the kernel length is equal to the input sequence length, and the convolution layer becomes a fully connected layer.

The second KINN layer nodes correspond to the KA diagram terms $\kappa_{\ell}$ in Equation \eqref{eq:ka}. The first and second layers are connected by a static(untrained) binary layer with the connectivity determined by the transitions in each KA diagram $\bB$. An example of such binary matrix representing KA diagram is shown in figure blah. 
The summation of rates in log-space implies a multiplication in linear-space. Therefore, given a sequence and applying a softmax activation, equation \eqref{eq:salpha} can be re-written as
\begin{equation}
\bs_\infty = \mathrm{softmax}(\bB \cdot \log(k\ab)).
\end{equation}

Finally, the effective cleavage rate is computed from equation \eqref{eq:act}. In our Cas9 massively parallel kinetic assay, this is the observed rate. To calculate activity, exponentiated log-rate nodes' values from the first layer multiply the enzymatic state before the activity-producing transition. If multiple rates contribute to activity, then all activity rates are included according to equation \eqref{eq:act}. A final activation function or mapping may be applied to the activity node if a measurement of the activity-measurement relation is known to be non-linear.
\begin{equation}
    \label{eq:y}
    y = \sigma(\act) = \sigma_o ( \sigma_\act(\act \cdot W_t + b_t) \cdot W_o + b_o ),
\end{equation}
where non-linear activation functions $\sigma_\act$, $\sigma_o$, and dimensions of $W_t$, can also be searched.


\section{Probabilistic model building genetic algorithm for searching KINN}
\label{sec:prob_model_search}
We employ a genetic algorithm to search for KINN model architectures. To incorporate existing knowledge and prior beliefs about a kinetic system, we use a Bayesian genetic algorithm based on probabilistic model building \cite{pelikan2011probabilistic}. For each kinetic rate constant, we first define a prior probability distribution for its hyperparameters: the sequence determinant $i,j$ and the convolution kernel size $d$. Then we sample model architectures and train the weights to evaluate performance. The search algorithm updates the posterior distribution by the architectures of each generation's surviving models.

Formally, let $\log(k_{\alpha\beta})=f(x_{i:j})$ be the log-scale kinetic rate that depends on subsequence from the $i$-th to the $j$-th nucleotide on the input sequence x; the function that maps $x_{i:j}$ to $\log(k_{\alpha\beta})$ is parameterized as a simple one-layer linear convolutional neural network.
As an uninformative prior, we set $i \sim$ Multinomial($[\phi(k)-w/2, \phi(k)+w/2]$), and $(j-i) \sim$  Multinomial([1,5]), where $\phi(k$) is the anchor function that uniformly maps each kinetic rate to sequence intervals on the input sequence x, and w is the search window size. For example, for a four-state {$s_0, s_1, s_2, s_3$} KINN with 20bp input sequence $\bx$, $\phi(k_{01})=5$, $\phi(k_{12})=10$, $\phi(k_{23})=15$.

In the first iteration, the genetic algorithm uniformly samples a number of m=10 KINNs from the prior probability distributions. The fitness of a KINN is defined as its performance on validation dataset, as measured by Pearson correlation between the prediction and observations. The expected fitness for each generation $t$, $C_t$, is evaluated by the average fitness values of KINNs sampled. Subsequently, KINNs with a fitness higher than $C_t$ survive and their model architectures are used to update the posterior distributions. 

\section{Benchmarking by simulated kinetic datasets}
\label{sec:benchmarking}
We tested our method of KINN optimization on simulated data where the kinetic rates' dependence gRNA-DNA alignment was known exactly. We started with an arbitrary sequence of 50 nucleotides. Only a subsequence from the 5th to the 40th nucleotide pairs contribute to the simulated cleavage rate. The 35 nucleotide pairings are separated into seven five-nucleotide long subsequences, each determining the value of a different kinetic rate constant for a 4-state Cas9 cleavage cycle. 

Transition rates where calculated from the Arrhenius equation
\begin{equation}
  \label{eq:forward_k}
  k\ab = k_o \exp \left[ \mp (k_bT)^{-1} \sum_{i\in\Omega\ab} \Delta \bar{G}\ab(x_i, \xi_i) \right]
\end{equation}
where the factor inside the exponent is negative for forward reactions ($\alpha < \beta$) and positive for backwards reactions ($\alpha > \beta$). $\Omega\ab$ is a set of pair indices $i$ that contribute to a reaction’s free energy barrier. No rates shared nucleotide pair dependence with other rates, i.e., $\Omega\ab \cap \Omega_{\alpha'\beta'}=\emptyset $ for $\alpha, \beta \neq \alpha', \beta'$.  $\Delta \bar{G}\ab(x_i, \xi_i)$ is the difference in free energy contributed by the $i$th substrate-gRNA nucleotide pair for the transition $\alpha \to \beta$. For this data, the values were taken from the set $\Delta \bar{G}\ab(x_i, \xi_i) \in \{-1, -0.1, 0.1, 1\}$. Complementary pairs were assigned a value of $-1$ but mismatched pairs varied for different transitions. All rates were multiplied by a baseline kinetic rate $k_o=.01$ sec$^{-1}$ to give the LHS the proper units.

Once the rate dependence was specified, 20000 sequences were randomly generated and the forward and backward rates calculated. From these rates, we constructed the rate matrix from equation (\ref{eq:bK}) 
\begin{equation}
  \label{eq:Kcas9}
  \bK_{cas9} = \left(
\begin{array}{cccc}
  -k_{01}D_u & k_{10} & 0 & k_{30} \\
  k_{01}D_u & -(k_{10}+k_{12}) & k_{21} & 0 \\
 0 & k_{12} & -(k_{21}+k_{23}) & k_{32} \\
 0 & 0 & k_{23} & -(k_{30}+k_{32})  \\
\end{array}
\right),
\end{equation}
where $D_u$ was the concentration of unbound DNA. The largest eigenvalue was deemed the cleavage rate for DNA-gRNA pairs, i.e., what KINNs were trained to predict.  

\section{Relation between King-Altman and single rate learning}%
\label{sub:relation_between_king_altman_and_single_rate_learning}
We have noticed that training our KA network on single rate data produces
accurate predictions on kinetic rates and activity despite being based on
different models.  This suggests an equivalency between the two ways to learn
$\bK(\bx)$. Here we show that under certain assumptions, the two methods are
approximately equivalent. First we construct a kinetic model with depleting DNA
\begin{equation}
  \label{eq:K'}
  \bK' = \left(
\begin{array}{cccc}
  -k_{01}D_u & k_{10} & 0 & 0  \\
  k_{01}D_u & -(k_{10}+k_{12}) & k_{21} & 0 \\
 0 & k_{12} & -(k_{21}+k_{23}) & k_{32}  \\
 0 & 0 & k_{23} & -(k_{34}+k_{32}) \\
\end{array}
\right)
\end{equation}
Notice the differences between equation (\ref{eq:Kcas9}) and (\ref{eq:K'}). The element $(\bK')_{3,3}=-(k_{34}+k_{32})$ while $(\bK_A')_{3,3}=-(k_{30}+k_{32})$, there are elements of $k_{01}D_u$ in $\bK_{cas9}$ instead of $k_{01}D_0$ as is in $\bK'$, and finally $(\bK')_{0,3} = 0$ while $(\bK_{cas9})_{0,3} = k_{30}$. The first difference is just a matter of notation since both rates produce the measurable quantity, cut DNA, and are therefore equivalent $k_{34} = k_{30} = k_{\rm cut}$. A similar argument can be made for the second difference if we were to switch $D_u$ and $S_0$. However, the final difference changes determinant of the matrices. Notice 
$$\det{\bK'}\neq0 \quad \text{while} \quad \det{\bK_{cas9}}=0$$ 
If we compare the characteristic polynomials $C(x;\bK)$ of these models 
$$C(x;\bK) = \sum_{n=0}^{N} a_nx^n = \prod_{n=1}^N(x - \lambda_n)$$
we see that $C(x;\bK')-C(x;\bK_{cas9}) = a_0({A}) = \det{\bK'}$. This is notable because $\det{\bK_A'} = k_{01}k_{12}k_{23}\kcut = \sum_{\alpha\beta \in {\rm act}} \tilde{k}\ab s_\alpha$ of system B. Even more interesting is that $a_1$ for both systems is the denominator of equation (\ref{eq:salpha}),  i.e. 
$$a_1(A) = \sum_{\ell'}\kappa_{\ell'}.$$
Therefore, small $|\lambda|$ values are approximately 
\begin{equation}
  \label{eq:lambda_approx}
  0 = C(\lambda;\bK'_A) = a_0 + a_1\lambda + O(\lambda^2)
\end{equation}
so 
\begin{equation}
  \label{eq:eig_ka_equiv}
  - \lambda \approx \frac{a_0}{a_1} = \frac{\sum_{\alpha\beta \in {\rm act}}\tilde{k}\ab s_\alpha}{\sum_{\ell'}\kappa_{\ell'}},
\end{equation}
which is the activity $\act$ given by the King-Altman equations (\ref{eq:salpha}) and (\ref{eq:act}). However, for more complicated systems, this approximation will not hold. Therefore, it might be necessary to apply a eigenvalue-solve layer instead of the KA layer.

\section{Convolutional neural network architecture search with pre-trained KINNs}
\label{sec:conv_pretrain_kinns}

We developed a new neural architecture search (NAS) method for automatically searching CNNs with pre-trained KINNs to model \textit{in vivo} kinetic systems. Let $\bar{\bk}=\{k_0, k_1, \ldots \}$ be a vector of log kinetic rates from a pretrained KINN \textit{in vivo}, and $\bk$ be the kinetic rates \textit{in vitro}, we seek to learn a modifier term $\delta_i$ for each kinetic rate $k_i$ such that $\delta_i$ account for the \textit{in vivo} sequence-context effects and $\bk = \bar{\bk} + \boldsymbol{\delta}$. 
These \textit{in vivo} kinetic rates $\bk$ are then fed into the downstream KINN layers of King-Altman layer and activity layer. The final binary output of cleavage probability is modeled by a sigmoid layer with a single scalar kinetic activity as input. 

We parameterized the functional dependency of $\boldsymbol{\delta}$ to the input gRNA and DNA sequences as a deep convolutional neural network backbone $f$. The input for $f$ is a 13-bit encoding for gRNA-DNA alignment by 25 nucleotides, where we used 4-bit to one-hot encode the gRNA sequence, 4-bit to one-hot encode DNA substitutions, 4-bit to one-hot encode DNA insertions, and 1-bit to encode DNA deletions. 
The 25 nucleotides consist of 3bp PAM site, 20bp gRNAs, and 2bp for padding DNA insertions.
The output from the CNN is a flatten vector, which will be linearly transformed to $\boldsymbol{\delta}$ and followed by $\bk = \bar{\bk} + \boldsymbol{\delta}$. 
The dimensionality of the linear transformation will be determined based on the pretrained KINN and its $\bar{\bk}$. 
The pretrained KINNs are converted to special layers with two inputs: one input directly from the gRNA-DNA alignments to compute $\bar{\bk}$, and the other input $\boldsymbol{\delta}$ from the deep CNN backbone.

To search for optimal CNN architecture, we employed AMBER (v0.1.3), an AutoML framework developed previously \cite{Zhang2021AMBER} and that demonstrated state-of-the-art NAS performance in genomics. AMBER searches neural network architectures with two components, 1) a controller model that optimizes neural architecture through reinforcement learning and 2) a model space to sample neural architectures. The search and optimization process of neural network architectures is described in detail previously \cite{Zhang2021AMBER}. 

Our model space for the CNN architectures consisted of 7 layers. All layer search spaces besides the first convolution layer, the flatten layer, or the KINN layer, include an identity operator which would effectively removes that layer if selected. This enabled the controller model to reduce model complexity during architecture optimization. The first two layers are convolution operators with different kernel sizes and dilation rates. Alternatively,  the controller can select specialized Inception layers that aim to capture patterns at multiple scales by combining kernels of different sizes into a single layer. Each convolution layer was followed by a search of dropout rates in dropout layers, with the goal of regularizing model complexity and improving model generalization. The 5th layer search space contained bidirectional LSTM layers with either 32 units or 16 units, with the option of an identity operator. This layer was followed layer with a search space that included a standard flattened layer, an attention-based flattened layer, a dense layer of 64 units, and an identity operator. Finally, we included the KINNs pretrained on the \textit{in vitro} data, with three different state numbers on two separate datasets, yielding a total of 6 pretrained KINN configurations.

\section{CRISPR/Cas9 Datasets and train-testing split}
\label{sec:train-test_split}

The datasets we used for creating the Elektrum model for CRISPR/Cas9 cleavage kinetics were divided into two categories. The first category consisted of time-resolved kinetic assays measured \textit{in vitro}. The second category consisted of \textit{in vivo} off-target editing measurement using high-throughput assays. We described the processing and train-test split within each category separately below.

For the \textit{in vitro} data, we re-compiled the Massively Parallel Kinetic Profiling (MPKP) for Cas9 generated previously \cite{Jones2021}. The MPKP measured the cleavage rate of SpCas9 on ~14,000 targets containing mismatches, insertions and deletions relative to two different gRNAs.  For each gRNA-DNA pair, we converted their sequences as described in the previous section to as a 13-bit by 25-bp encoding matrix. For each sgRNA, we trained a separate model using all its data points, while used 50\% of the other sgRNA datapoints as validation data to search for neural network architectures, and the remaining 50\% of the other sgRNA data points as a held-out testing data.

For the \textit{in vivo} data, we used the assembled training dataset from CRISPR-Net \cite{lin2020crispr}. Briefly, this training data was collected and uniformly processed from 8 datasets in previously published Cas9 off-target assays \cite{tsai2017circle, listgarten2018prediction, doench2016optimized, haeussler2016evaluation, cameron2017mapping, kleinstiver2015engineered}. In this data, the negative data points (that is, inactive off-target sites) were constructed by searching genome-wide DNAs by Cas-Offinder \cite{bae2014cas}, a versatile tool that searches for potential off-target sites of Cas9. To ensure a fair comparison, we held out the same three testing datasets to match the original train-test split as in CRISPR-Net. In total, the testing dataset has 693,235 gRNA-DNA pairs. For the remaining data, we randomly split the dataset by gRNA sequences to hold-out 20\% of the gRNA and all their corresponding gRNA-DNA pairs as validation datasets. This corresponds to a total of 842,298 gRNA-DNA pairs for training, and 269,900 gRNA-DNA pairs for validation.

\section{Neural architecture analysis for KINN and Elektrum}
\label{sec:arch_anal_elektrum}

To evaluate the neural architectures from KINN search on the Cas9 MPKP and account for stochasticity between different search runs, we developed a Baysian hierarchical model to infer the maximum \textit{a posteriori} estimate for each neural architecture parameter. For example, the rate sequence anchor point and its sequence determinant length were likely co-varied. The Baysian hierarchical model allowed us to consider these uncertainty and covariance in a principled framework. Specifically, we employed a multivariate Normal (MVN) distribution for the group-level architecture parameters whose mean follows an uninformative prior and variance-covariance matrix is defined a distribution over Cholesky decomposed covariance matrices, such that the underlying correlation matrices follow an LKJ distribution \cite{lewandowski2009generating} with the standard deviations $\eta=1$. We assumed each replicate run draws a set of architecture parameters from the group-level MVN distribution, such that the optimized KINNs follows another within-group MVN with its mean determined by the group-level sampled parameters. The covariance matrix of the within-group MVN followed an uninformative prior of an LKJ distribution with the standard deviations $\eta=1$. The top 5\% architectures from each run were fed into the Baysian model for parameter fitting. We implemented the above model using PyMC3 \cite{Salvatier2016}.

For the Elektrum selection probability, because each layer was modeled by a recurrent neural network for the dependency from its preceding layers without complex multivariate dependency, we simply pooled the last five architectures from each search run across five independent search runs. Then we computed the mean and variance for the likelihood of selecting a candidate layer within the given model space.

\section{Base-pair importance interpretation by SHAP}
\label{sec:shap_interp}

In order to determine the base-pairs that are important in explaining the difference between \textit{in vitro} and \textit{in vivo} cleavage predictions, we applied the SHAP analysis to both the base KINN and the trained Elektrum model. Both KINN and Elektrum were end-to-end differentiable, which allowed us to use the GradientExplainer to assess the importance of each base-pair within each model. Gradient explainer seeks to explain a model using expected gradients, which is an extension of integrated gradients \cite{NIPS2017_7062}. We then normalized the maximum absolute base-pair importance to 1, making the importance scores comparable across different predictions. After obtaining the normalized base-pair importance scores for both \textit{in vitro} and \textit{in vivo} cleavage predictions, we subtracted the \textit{in vitro} scores from the \textit{in vivo} scores to determine the specific importance of each base-pair in explaining the difference between \textit{in vitro} and \textit{in vivo} cleavage. In this final base-pair importance matrix, if a base-pair contribution is identical \textit{in vivo} vs \textit{in vitro}, it’s score would be zero; in contrast, a positive importance score indicates the activation of cleavage \textit{in vivo} specifically, and a negative value indicates the inactivation \textit{in vivo} specifically. To compute the integrated gradient values, we used an all-zero matrix of shape 25 by 13 as the background. Finally, using the five models constructed by independent AMBER runs, our interpretation method of base-pair effects considers both the averaged Shap values as well as its variance across models. 

\section{Data Availability}
\label{sec:data_avail}
All training and testing data presented in this paper are publicly available. The \textit{in vitro} kinetic data was downloaded from Jones et al. \cite{Jones2021} in URL \url{https://github.com/finkelsteinlab/nucleaseq/tree/master}. The \textit{in vivo} CIRSPR/Cas9 off-targeting dataset was downloaded from CRISPR-Net \cite{lin2020crispr} in URL \url{https://codeocean.com/capsule/9553651/tree/v2}.

\section{Code Availability}
\label{sec:code_avail}
The \textit{Elektrum} code is available on GitHub at \url{https://github.com/zj-zhang/Elektrum}.

\bibliography{elektrum, added}